\documentclass[prl,aps,10pt,twocolumn]{revtex4-1}
\pdfoutput=1

\usepackage{color}
\usepackage{slashed}
\usepackage{hyperref}
\usepackage{amsmath}
\usepackage{amssymb}
\usepackage{latexsym}
\usepackage{amsfonts}
\usepackage{epsfig}
\usepackage{psfrag}
\usepackage{graphicx}
\usepackage{mathtools}
\usepackage{array}

\def\hhref#1{\href{http://arxiv.org/abs/#1}{arXiv:#1}} 

\newcommand{\bea}{\begin{eqnarray}}
\newcommand{\ea}{\end{eqnarray}}
\newcommand{\eea}{\end{eqnarray}}

\DeclareMathOperator{\sech}{sech}

\begin{document}

\title{Multidimensional Quantum Tunneling in the Schwinger Effect}
\author{Cesim K. Dumlu}
\affiliation{Department of Physics, Middle East Technical University, 06800, Ankara, Turkey}
\email{cdumlu@metu.edu.tr}

\begin{abstract}
We study the Schwinger effect, in which the external field having a spatiotemporal profile creates electron-positron pairs via multidimensional quantum tunneling. Our treatment is based on the trace formula for the QED effective action, whose imaginary part is represented by a sum over complex worldline solutions. The worldlines  are multiperiodic, and the periods of motion collectively depend on the strength of spatial and temporal inhomogeneity. We argue that the classical action that leads to the correct tunneling amplitude must take into account both the full period,  $\tilde{T}$ and the first fundamental period, $T_1$. In view of this argument we investigate pair production in an exponentially damped sinusoidal field and find that the initial momenta for multiperiodic trajectories lie on parabolic curves, such that on each curve the ratio $\tilde{T}/T_1$ stays uniform. Evaluation of the tunneling amplitude using these trajectories shows that  vacuum decay rate is reduced by an order of magnitude, with respect to the purely time-dependent case, due to the presence of magnetic field.

\end{abstract}



\maketitle

The surge of interest on the Schwinger effect, the nonperturbative production of electron-positron pairs from vacuum in an external electric field, 
has yielded new insights into this peculiar yet unobserved prediction of QED \cite {he, schw}.Going beyond uniform field approximation, computations of vacuum decay rate show that mean number of produced particle pairs depends nontrivially on the shape, cycle structure and polarization of the external pulse \cite{nazo1,dunne1,gies1,pizza1, dumlu3}. While such investigations predominantly deal with the time-dependent electric fields, laboratory fields are usually composed of Gaussian or x-ray beams, which have spatiotemporal profile and include magnetic fields as well. This makes the formulation of vacuum decay in multidimensional electromagnetic fields essential for 
the fully realistic treatment of the problem.

The technical challenge is to compute the imaginary part of QED effective action, $\text{Im}\,\Gamma[A_{\mu}]$, which 
requires the knowledge of vacuum persistence amplitude in the background gauge field,  $A_{\mu}(x)$.  The standard approach is to use the S-matrix formalism, which  relates  $\text{Im}\,\Gamma[A_{\mu}]$  to the WKB coefficients of the vacuum state by a Bogoliubov transformation\cite{popov1, kluger1}.   While this approach works  perfectly well for one dimensional external fields, its extension to higher dimensional,  nonseparable backgrounds is exceedingly difficult.  One way to circumvent this difficulty is the  inverse scattering method, where one starts with a plausible ansatz for the Dirac equation to establish a physical set of gauge configurations\cite{schutzhold}. A more direct approach involves brute force  integration of Dirac equation on a spatial grid.  An  application of this to  Schwinger effect was given in \cite{ruf}.  In connection with the standard scattering methods, works aiming at generalization of quantum kinetic equation to 1+1 dimensions can be found in \cite{hebenstreit1}. Apart from the technical subtleties of the chosen method,  the integrability  of multidimensional  Dirac Hamiltonian becomes an important aspect of the problem, because it relates  to the question of what the conserved quantities are in a given background. Apparent translational symmetry of the Hamiltonian in one dimensional backgrounds makes the quantization of the created particle pairs by the conserved momentum (or energy)  straightforward. In the multidimensional setting, identification of particle content remains elusive within the framework of Bogoliubov-type transformations. Evidently, these problems also appear in semiclassical WKB analysis, which uses analytic continuation of WKB solutions in the complex domain\cite{dumlu1,kim1}.

Here, we give semiclassical treatment of the multidimensional vacuum pair production by using the worldline formulation of QED  \cite{rubakov1, dunne2, ilderton}. The worldline approach could be considered more advantageous with respect to conventional WKB methods for two  reasons. First reason is that in the worldline language  $\text{Im}\,\Gamma[A_{\mu}]$ is represented by a path integral  over closed trajectories in space-time, thus the formalism admits a natural multidimensional  description. Secondly, no specific choice of ansatz and Bogoliubov transformation are needed;  calculation of  $\text{Im}\,\Gamma[A_{\mu}]$ is relegated to finding periodic, tunneling trajectories, which are also referred as worldline instantons. Basic formalism is not just relevant to Schwinger effect, but it also has the potential to deal with multidimensional tunneling problems in general.

Main ingredient of the following analysis is the  QED analog of widely used trace formula for the Green function\cite{gutzwiller1, dietrich1}. The trace formula can be obtained upon performing saddle point approximation to $\Gamma[A_{\mu}]$, which can be written as a sum over closed orbits  ($ g_{\mu\nu}= (+,-,-,-,) ,\,\,\hbar=c=1$):
\begin{eqnarray}
&&\Gamma[A_{\mu}] \approx  -\frac{i}{2}\sum_{\substack{\text{$p$}}}
 \frac{e^{-i W[T_p] - i m_{p} \pi/2}}{\sqrt{{\text{det}\,\bf{\Lambda}} }}\,  {\rm tr}\left[e^{-i \frac{|e|}{4} \int_0^{T_p} \sigma^{\mu\nu}F_{\mu\nu} du}\right]
 \nonumber\\
&&W[T_p]=\int_{0}^{T_p} p_{\mu} \dot{x}^{\mu} du,\quad \sigma^{\mu\nu}= \frac{i}{2}\left[\gamma^{\mu},\, \gamma^{\nu}\right]
\label{seffact}
\end{eqnarray}
The matrix  $\bf{\Lambda}$ comes from the saddle point expansion and corresponds to the density of the trajectories which start at the same point with different initial momenta. For each periodic trajectory labeled by $p$, the Morse index, $m_p$ is given by the number of negative eigenvalues of the determinant, whereas the spinor term contributes a $+/-$ sign. The determinant can analytically  be obtained when equations of motion separate, and the  trajectory admits a simple form. Here, we will work with nonseparable cases and focus our attention to the classical action, $W[T]$, which is sometimes referred as the Hamilton's characteristic function. Semiclassical value of $W[T]$ is given by the classical tunneling trajectories, which are imaginary proper time solutions of the force equation:
\begin{eqnarray}
\ddot{x}_{\mu}=i|e|\,F_{\mu\nu}(x) \dot{x}^{\nu}, \qquad  u \rightarrow iu
\label{class}
\end{eqnarray}
Periodic solutions of (\ref{class}) also exist for real $u$ \cite{dumlu2}. Composite worldlines which are made of $x(u)$ and $x(i u)$ are responsible for possible interference effects, but their contribution to pair production rate is controlled by the tunneling segments $x(i u)$.  For this reason, we are interested only on imaginary proper time solutions, as they are expected to yield a pretty accurate estimate in the nonperturbative domain. In the following, we argue that worldline trajectories are  multiperiodic in spatiotemporal backgrounds. Starting with a simple model, we discuss the implications of bounded motion in the multidimensional setting, and move onto a more realistic scenario incorporating the magnetic field.

\begin{figure}[htb]
\includegraphics[width=4.25cm,height=2.7cm]{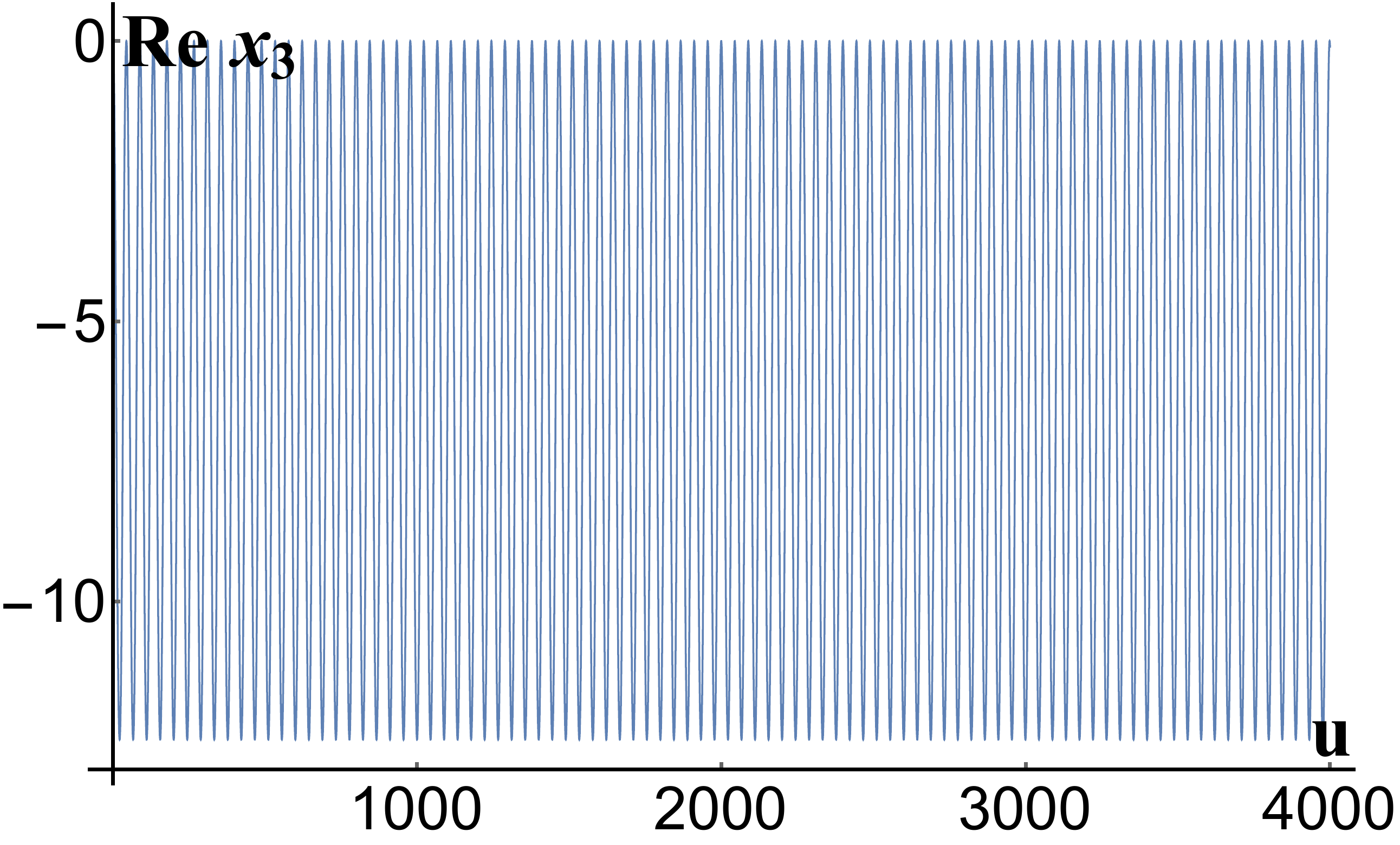}
\includegraphics[width=4.25cm,height=2.7cm]{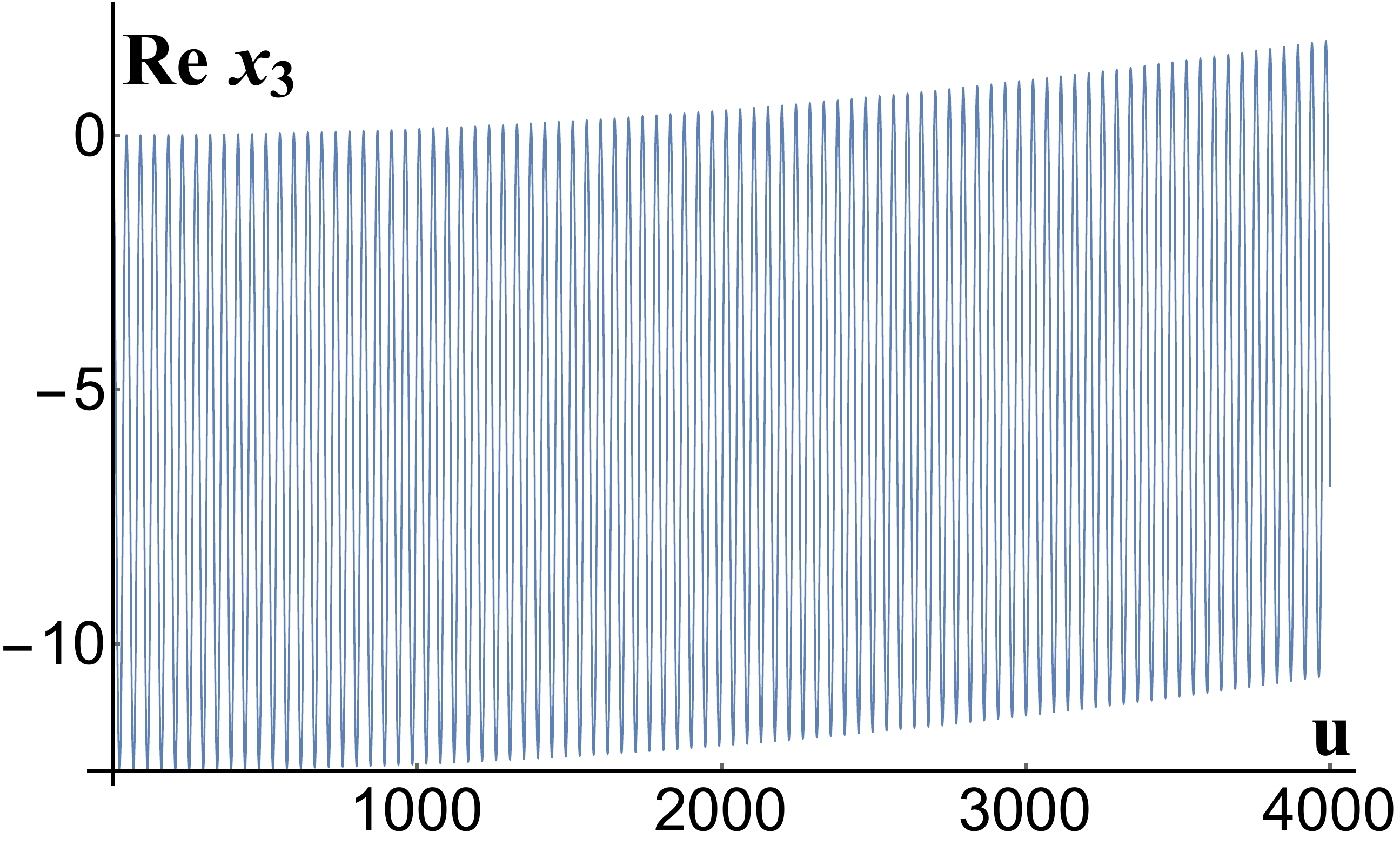}
\includegraphics[width=4.2cm,height=2.7cm]{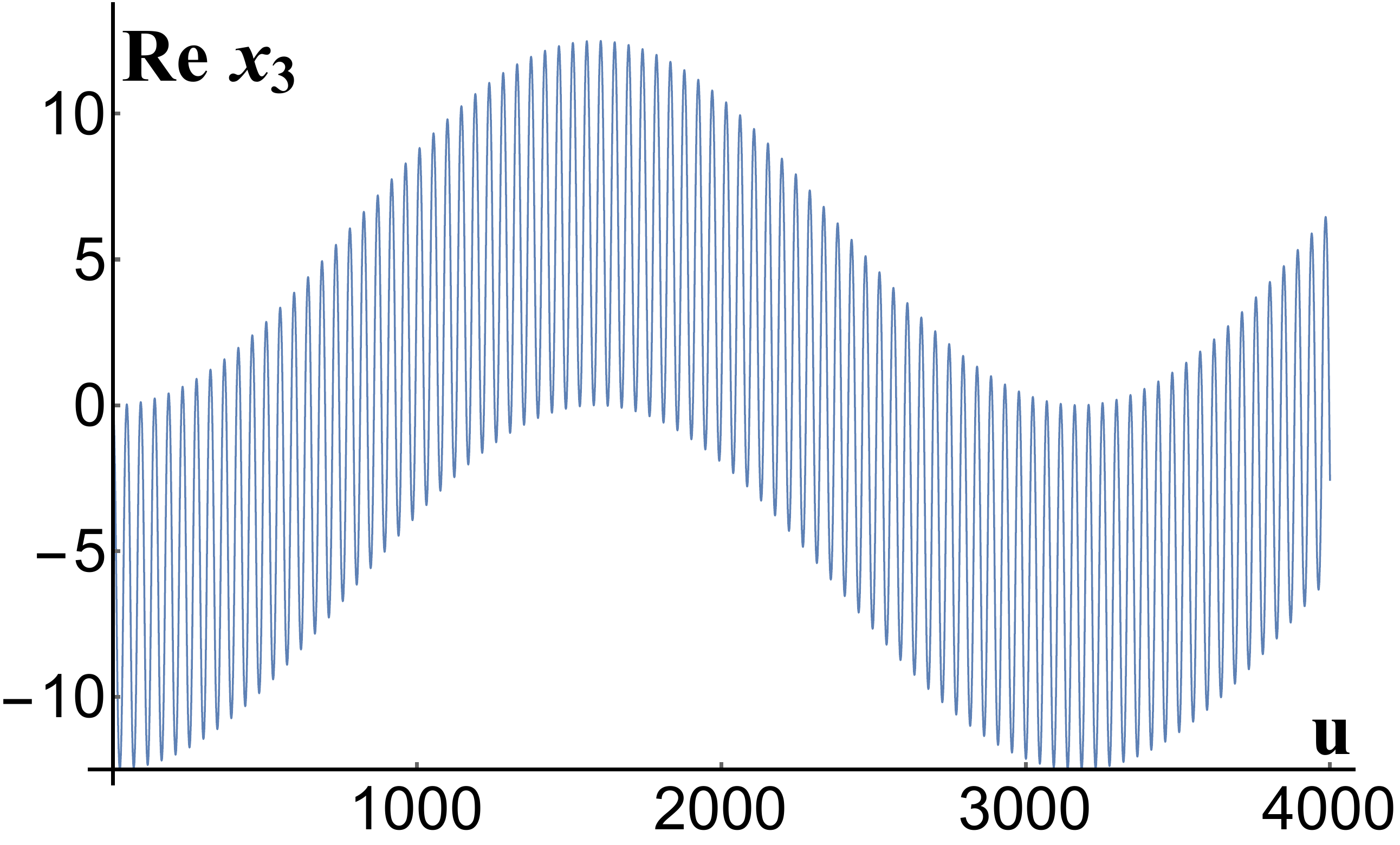}
\includegraphics[width=4.2cm,height=2.7cm]{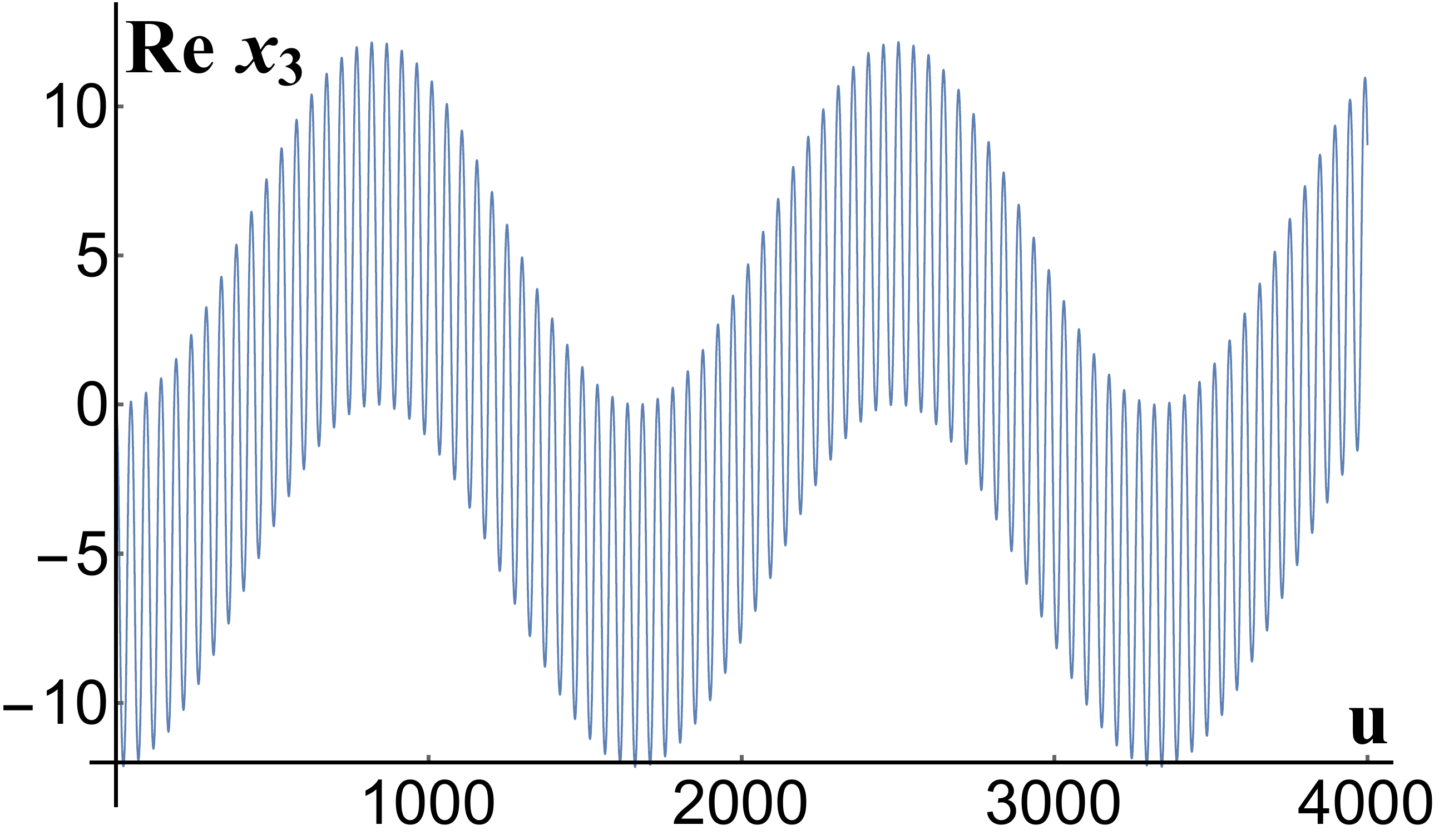}
\caption{Worldline trajectories in the background field (\ref{saut}), as a function of propertime $u$.  Temporal parameters are fixed as $E_0= 0.1\, m$, $\omega= 0.1\, m$. Spatial parameters are chosen as $E'_0=0 $  (top-left),    $E'_0=0.01\, m^2$, $k=0.001\, m$ (top-right)  $E'_0=0.01\, m^2$, $k=0.01\, m$ (bottom-left), and  $E_0'=0.03\,m^2$, $k=0.01\, m$  (bottom-right) }
\label{f1}
\end{figure}

We first consider the combination of two pulse-shaped electric fields. Total field  along $x_3$ is given as:
\begin{eqnarray}
A_{x_0}(x_3) &=& \frac{E_0}{\omega} \tanh{\omega x_0}, \quad A_{x_3}(x_0)= \frac{E'_0}{k} \tanh{k x_3}\nonumber\\
E_{x_3}(x_0,\, x_3)&=& E_0 \sech^2{ \omega x_0} + E_0' \sech^2{ k x_3}
\label{saut}
\end{eqnarray}
where $k$ and $\omega$ represent spatial and temporal width respectively. To illustrate the effect of spatial inhomogeneity on the closed orbits, we keep the temporal parameters fixed, and vary $E'_0$ and $k$. As observed in \cite{dumlu2}, in the purely time-dependent case ($E'_0=0$) tunneling trajectories are periodic with the classical period being $T$, and quantized by the canonical momentum $p_3$. Introduction of spatial inhomogeneity causes periodic trajectories to have a second oscillation period $\tilde{T}$, which envelopes the oscillations with a smaller period, $T_1$ (Fig. \ref{f1}).  These trajectories are quasiperiodic and form invariant tori in phase space. Given the time component of the above field is dominant ($E'_0 < E_0$), such bounded trajectories generally persist for a finite range of momenta, when $k < \omega$.  Our key observation is that in the constant spatial field limit: $k \rightarrow 0$, and also in the limit: $E'_0 \rightarrow 0$, $\tilde{T}$ increases and effectively goes to infinity, leaving a single period $T_1\equiv T$. This indicates finite values of $T_1$  and $\tilde{T}$ genuinely depend on the interplay between the temporal and spatial adiabaticity parameters of the external field. This applies not just to this particular case but holds general validity.  Few other examples include  pulse configurations such as:
\begin{eqnarray}
E_{x_1}&=&E_0\,e^{-k^2 x^2_3 - \omega^2 x^2_0},\,\,E_{x_1}= E_0\, e^{-k^2 x^2_3} \sech{ \omega x_0}^2, \nonumber\\ 
E_{x_3}&=&E_0\, e^{-\omega^2 x^2_0} + E'_0\,e^{-k^2 x^2_3}\nonumber
\label{ex}
\end{eqnarray}
Note that Hamiltonian, $\mathcal{H}=1/2 \left(p_{\mu}-e/c\, A_{\mu}(x)\right)^2$,  for the external field in (\ref{saut}) has no longer translational symmetry along $x_3$. On the other hand, existence of quasiperiodic motion tells us that system possesses dynamical (hidden) symmetry. This is because Poincar\'e sections of the quasiperiodic orbits form closed curves in the reduced phase space, and imply the existence of an integral of motion $\mathcal{C}$, in addition to $\mathcal{H}$. One may hope in this case the corresponding quantum system possesses simultaneous eigenstates of constants of motion by the virtue of Liouville integrability. Quantum tunneling via multiperiodic motion can then be interpreted as the creation of particle states quantized by $\mathcal{\hat{H}}$ and $\mathcal{\hat{C}}$.

The tunneling amplitude for the vacuum decay is governed by the action, whose evaluation on multiperiodic solutions must be handled with care. Direct approach would involve computing $W[\tilde{T}]$, where $T$ simply gets replaced by $\tilde{T}$. This however  leads to an undesirable limiting behavior for the tunneling amplitude. To explain we reconsider the toy model above. For a weak spatial inhomogeneity ($E'_0 \ll E_0$) we have the second period extended over a very large proper time scale compared to $T_1$, so we have $T_1 \ll \tilde{T}$. In this regime, the smaller period almost coincides with the period of purely time-dependent background, whose action is denoted by $W[T]$.  On the other hand, the evaluation of $W[\tilde{T}]$ yields $\sim \tilde{T}/T_1\, W[T]$. Thus the corresponding vacuum decay rate is many orders of magnitude smaller than $e^{- W[T]}$, which gives the decay rate for the time-dependent field. Such a huge difference in tunneling probabilities is quite unexpected for a weak spatial perturbation. In fact, in the limit: $E'_0\rightarrow 0$, $\tilde{T}$ gets infinitely large and so $W[\tilde{T}]$ will get infinitely large. This basically leads to vanishing pair production probability, which is clearly in contradiction with the purely time-dependent case. To remedy this situation tunneling treatment of the trajectories should take the internal oscillation period
$T_1$ into account. To incorporate $T_1$ into this picture, we rewrite the effective action as:
\begin{align}
\Gamma[A]
&= \frac{i}{2}\text{tr\,Log} \left[ m^2 + \slashed{D}^2 \right] \nonumber\\
&= -\frac{i}{2}\int_0^\infty \frac{dT_1}{T_1} \,{\rm tr}\,\text{exp}\left[-\frac{i T_1}{2 \tilde{T}} (m^2 + \slashed{D}^2)\, \tilde{T}\right] \nonumber\\
&\approx  -\frac{i}{2}\sum_{\substack{\text{$p$}}}\, e^{-i W[T_1, \,\tilde{T}]} \, {\rm tr}\left[e^{ -i \frac{|e|T_1}{4 \tilde{T}} \int_0^{\tilde{T}} \sigma^{\mu\nu}F_{\mu\nu} du}\right], \nonumber\\
\slashed{D}=&  \, \gamma^{\mu} (\partial_{\mu}- i e A_{\mu}),  \,\,
W[T_1,\,\tilde{T}]=\frac{T_1}{\tilde{T}}\int_{0}^{\tilde{T}} \hspace*{0cm}p_{\mu} \,\dot{x}^{\mu} du
\label{eff2}
\end{align}
Integration variable in the definition of the logarithm above was chosen to be $T_1$, whose appearance would produce the desired limiting behavior for the tunneling amplitude. Through redefinition of the exponent in second line of \ref{eff2} and fixing the upper limit of the ordering parameter u as $\tilde{T}$, the trace operation is to be performed over the space of multiperiodic trajectories, for which $T_1/ \tilde{T}$ is a constant factor. With this in mind and making the change of variable ($T_1 \rightarrow  \tilde{T}$) in the integration, saddle point approximation over $\tilde{T}$ can be carried out in the usual way. At first look the transformation of the exponent in \ref{eff2}  is very similar to gauge fixing of the effective action; but it does not correspond to a simple deformation of the period, $\tilde{T}$, rather it represents $W [T_1, \,\tilde {T}]$ as an average over the internal cycles of the trajectory. Note that the evaluation of $ W[T_1, \,\tilde {T}]$ on quasiperiodic solutions not so surprisingly yields a small, unphysical imaginary part, because the trajectory does not close back on itself at some finite $\tilde{T}$. Yet this imaginary part tends to vanish in the limit $\tilde{T} \rightarrow \infty$, which precisely corresponds to taking the average of a quasiperiodic function. The question of whether such quasiperiodic averaging may contribute, or assist to quantum tunneling might be interesting in its own right. In the remainder of this work however, we will focus on the closed orbits.

\begin{figure}[t!]
\includegraphics[width=4.2cm,height=4cm]{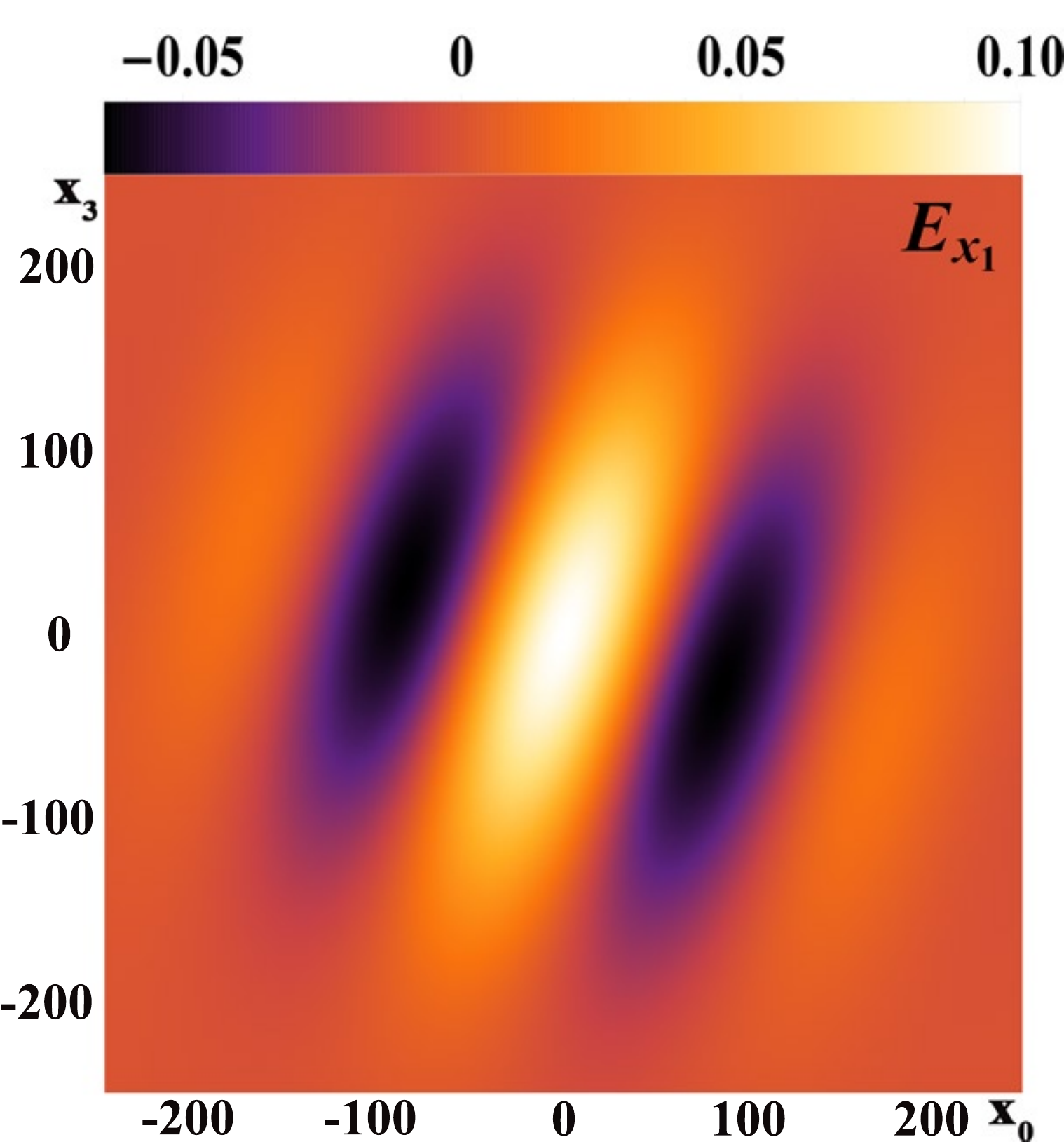}
\includegraphics[width=4.2cm,height=4cm]{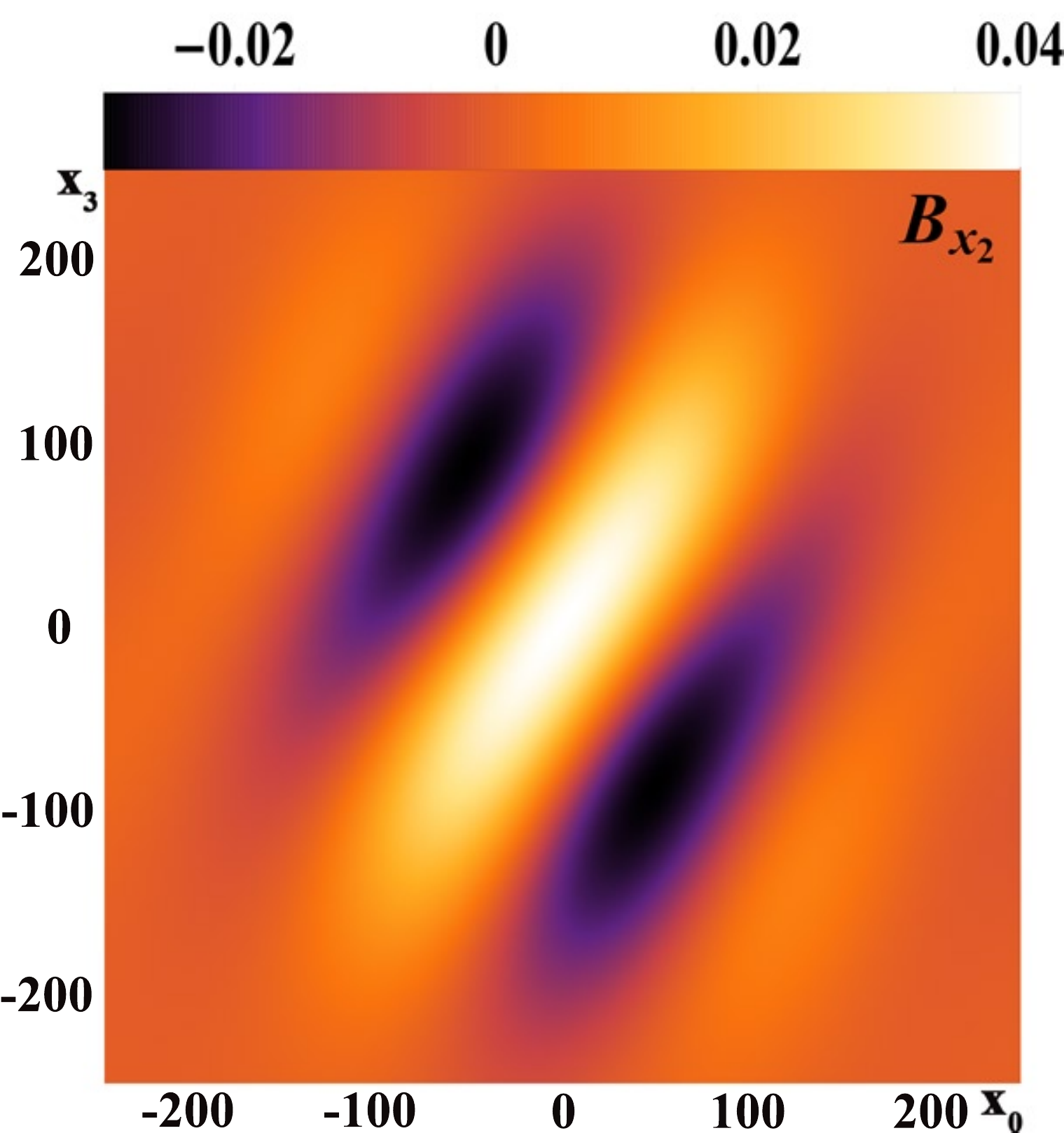}
\caption{External fields represented by (\ref{field}). Field parameters are given in the text. }
\label{f2}
\end{figure}

In the following we introduce spatial variation to the electric field: $E_1(x_0)=  E_0\, e^{-x_0^2/2 \tau^2}\hspace*{-.15cm}\cos{\omega x_0}$, whose corresponding vacuum decay rate was analyzed in \cite{gies1}. This field configuration approximates the experimental setup where two counter-propagating short laser beams form a standing wave. Interaction region is conveniently chosen along the beam, on the plane where magnetic fields cancel. Here, we release this restriction by taking into account the  spatial profile of the external field, along the beam direction, $x_3$. For this, we consider the transverse field:
\begin{eqnarray}
A_{x_1}(x_0,\,x_3)&=&-E_0 e^{-\frac{x_3^2}{2 \sigma^2}-\frac{ \tau^2\omega^2}{2}}\left(e^{-i k x_3}\,f(x_0)+ \text{c.c}\right),\nonumber\\
E_{x_1}(x_0,\,x_3)&=&  E_0 e^{-\frac{x_0^2}{2 \tau^2}-\frac{x_3^2}{2 \sigma^2}}  \cos{\left(\omega\, x_0 - k\, x_3\right)},\nonumber\\
B_{x_2}(x_0,\,x_3)&=&  E_0 e^{-\frac{x_3^2}{2 \sigma^2}-\frac{ \tau^2\omega^2}{2}}\left(e^{-i k x_3}\, g(x_3) f(x_0)+ \text{c.c}\right),\nonumber\\
f(x_0)&=&\frac{\sqrt{\pi}\tau\text{Erf}\left[\frac{x_0-i\tau^2\omega}{\sqrt{2}\tau}\right]}{2\sqrt{2}},\,\,\, g(x_3)=\frac{x_3}{ \sigma^2}+ik
\label{field}
\end{eqnarray}
with the frequency, $\omega$ and the wavenumber, $k$ (Fig. \ref{f2}). The temporal and spatial width are respectively given by $\tau$ and $\sigma$. The exponential terms above make the external field finite, whereas oscillatory terms  simulate the intensity variations typically seen in the Gaussian or x-ray pulses.  Note that above field introduces a source current along $x_1$. This current  may in principle affect the tunneling amplitude, because it contributes to the evolution of the momentum operator, $d\hat{x}_{1}/du$. However, the effect of the source term remains negligible, and classical solutions are expected to dominate tunneling, as long as the external field extends over a distance larger than the Compton wavelength\cite{note}.

\begin{figure}[b!]
\includegraphics[width=6cm,height=3.85cm]{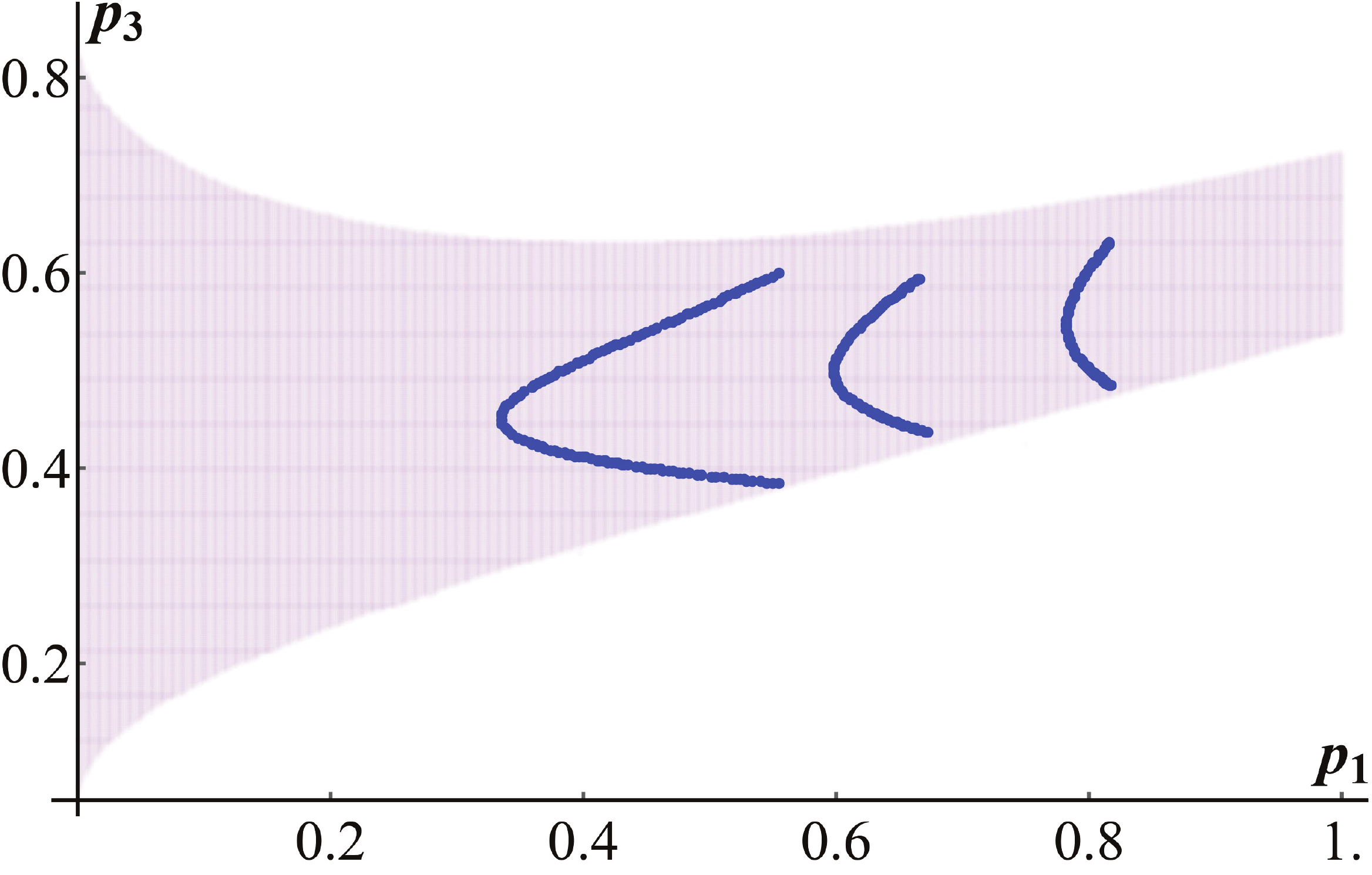}
\includegraphics[width=6.2cm,height=3.85cm]{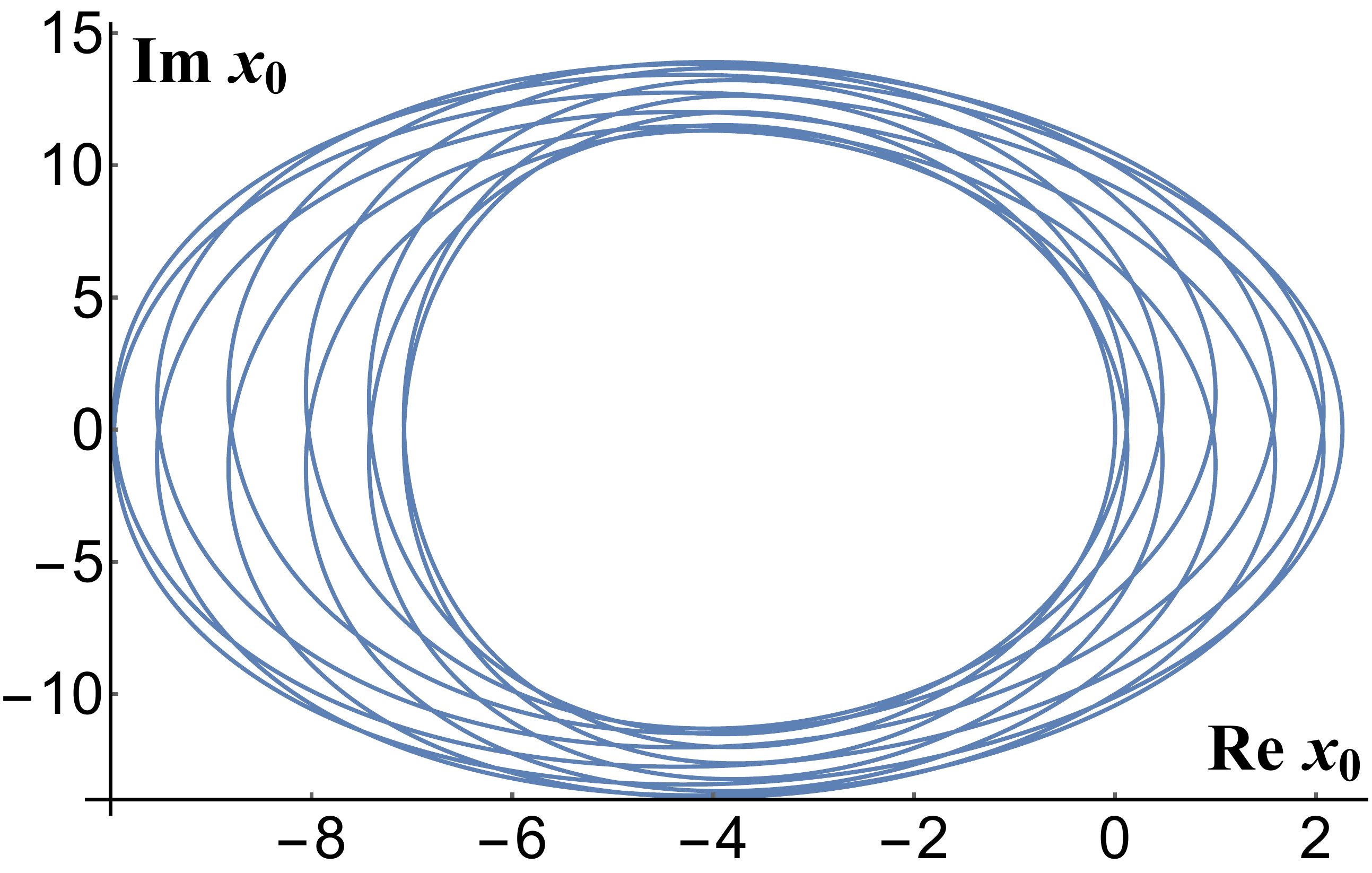}
\caption{Location of the quasiperiodic trajectories on $p_1-p_3$ plane (shaded). Multiperiodic trajectories are located on the curves (top). The figure below shows  a typical closed trajectory with  $p_1= 0.35351$ and $p_3= 0.42802$}
\label{f3}
\end{figure}

The existence of invariant tori now depends on the relative magnitude of both $\sigma$ and $k$ with respect to the temporal parameters.  Quasiperiodic trajectories generally occur in the parameter region where $k\sigma < \omega\tau$ and  $k < \omega$.  Here, we look for the closed orbits when the electric field is maximum, so we fix the initial positions as $x_0(0)=x_3(0)=0$, and vary the conserved momentum $\dot{x}_1[0]=- i p_1$, and the initial velocity $\dot{x}_3(0)\equiv -i p_3$. The remaining initial condition for $\dot{x}_0(0)\equiv-i p_0$ is fixed by the constraint: $\dot{x}_0^2(u)-\dot{x}_1^2(u)-\dot{x}_3^2(u)=-m^2$, where $m$ is the electron mass.  For the sake of comparison, we specify the values of $E_0$, $\tau$ and $\omega$, in accordance with \cite{gies1}. In terms of the normalized mass ($m=1$) we fix the field parameters as: $E_0=0.1\, m^2, \tau=100\, m^{-1},\, \omega=0.03\, m$. To see the full effect of spatial dependence and the arising magnetic field, we will work in the parameter region where temporal and spatial inhomogeneities become equally important. We accordingly choose the spatial frequency to be $k=0.01 m$ and $\sigma=100\, m^{-1}$. Multiperiodic orbits are located by making use of a search algorithm that scans through the region: $p_1 \in (0 , m), \, p_3 \in (0 , m) $, until the  trajectory closes on itself within an accuracy of $10^{-6}$. The greater accuracy makes the location of the orbits more precise but, from a practical point of view,  increasing the accuracy further does not have appreciable effect on the tunneling rate.  Figure \ref{f3} shows the locations of the closed orbits on the momentum plane; multiperiodic trajectories are not isolated, but form a family. Along each parabolic curve representing the orbit family, trajectories grow in amplitude and get steeper until the invariant tori break and the motion becomes unbounded. For every closed trajectory on the first parabolic curve we have $\tilde{T}/T_1=12$, on the second, period doubling occurs and ratio becomes 23, and on the third curve it becomes 11. On the first and third curve $\tilde{T}$ matches with the second fundamental period, $T_2$, whereas on the second curve we have $T_2 = \tilde{T}/2$. Accordingly, the fundamental frequencies of the system consists of set of co-prime integers, which are respectively given as: $(12, 1),\, (23, 2),\, (11, 1)$. These sets of integers characterize the orbit topology\cite{berry1}. Consequently, \ref{seffact} is considered as a topological sum, where each curve family is labeled by its topological index.

Before we finally evaluate  $W[T_1, \,\tilde {T}]$, we analyze the spinor term, which leads to an interesting geometric interpretation for the pair production process. Straightforward calculation of the spinor trace for the field configuration (\ref{field}) gives ($u\rightarrow i u$)
\begin{eqnarray}
&& {\rm tr}\left[e^{- \frac{T_1}{4\tilde{T}}\int^{\tilde{T}}_0 \ \sigma^{\mu\nu}(i |e|F_{\mu\nu}) du}\right] = \cos{\left(\frac{T_1}{2 \tilde{T}}\sqrt{\mathcal{E}^2-\mathcal{B}^2}\right)}\nonumber\\
&&\mathcal{E}=|e|\int_0^{\tilde{T}}\hspace*{-.3cm} E_{x_1}\,du, \quad \mathcal{B}=|e|\int_{0}^{\tilde{T}} \hspace*{-.3cm} B_{x_2} \, du
\label{spi}
\end{eqnarray}
The exponent in \ref{spi} has the form of a Lorentz transformation: as the spin  precesses under the influence of external field, the amplitude $e^{-W[T_1, \,\tilde {T}]}$ goes under a rotation in the complex plane given by the angle $T_1/4\tilde{T}\int_0^{\tilde{T}} \hspace*{-.15cm} F_{\mu\nu}\, du $. It can readily be shown by integration that for closed trajectories the value of $\sqrt{\mathcal{E}^2-\mathcal{B}^2}$ is precisely $2\pi \, \tilde{T}/T_1$.  Geometric meaning of this result becomes more transparent if one considers the relation between $\sqrt{\mathcal{E}^2-\mathcal{B}^2}$ and the curve invariants of the motion with the aid of Frenet-Serret formulas. For instance, in the case of single dimensional inhomogeneities geodesic curvature of a closed orbit is given by $|e| E_{x_1}$. Consequently, the argument $\mathcal{E}$ is nothing but the total curvature, $\mathcal{K}$, whose value is $2\pi$, regardless of the shape of the electric field. This is because instanton trajectories with 1 degree of freedom are simple closed loops, which can be continuously deformed to a circle.  With 2 degrees of freedom trajectories become non-planar, and geodesic curvature can not be solely given in terms of the field strength. Also, total torsion along the closed orbit may not vanish, unless the given trajectory is symmetric under reflections with respect to the normal plane at $u=0$. In the general case the relation between the field strength and the curve invariants can be given by:
\begin{eqnarray}
 \lVert e\hspace*{-.12cm}\int_0^{\tilde{T}} \hspace*{-.15cm} F_{\mu\nu}   \,du\, \rVert \approx \sqrt{\mathcal{K}^2-\mathcal{T}^2}
\label{spicur}
\end{eqnarray}
The norm above gives the total rotation angle in inhomogeneous fields, whereas the right-hand side could be regarded as the total invariant curvature of the trajectory.  For the closed curves considered here $\mathcal{K}$ is very close to $2\pi \,\tilde{T}/T_1$, whereas the total torsion, $\mathcal{T}$, vanishes. 
For slowly varying transverse electromagnetic fields, it could be said  without loss of generality that the imaginary part of $\Gamma[A_{\mu}]$ follows from the quantization of the rotation angle by the integer multiples of $2\pi$. But unlike the uniform or purely time-dependent cases, quantization condition in spatiotemporal backgrounds  depends on the set of values that $\tilde{T}/T_1$ can take. If the semiclassical approximation holds well,  it is then evident that the existence of  worldline trajectories with quantized invariant curvature is intimately connected to the nonperturbative pair production.

\begin{figure*}[htb!]
\includegraphics[width=5.9 cm,height=3.1cm]{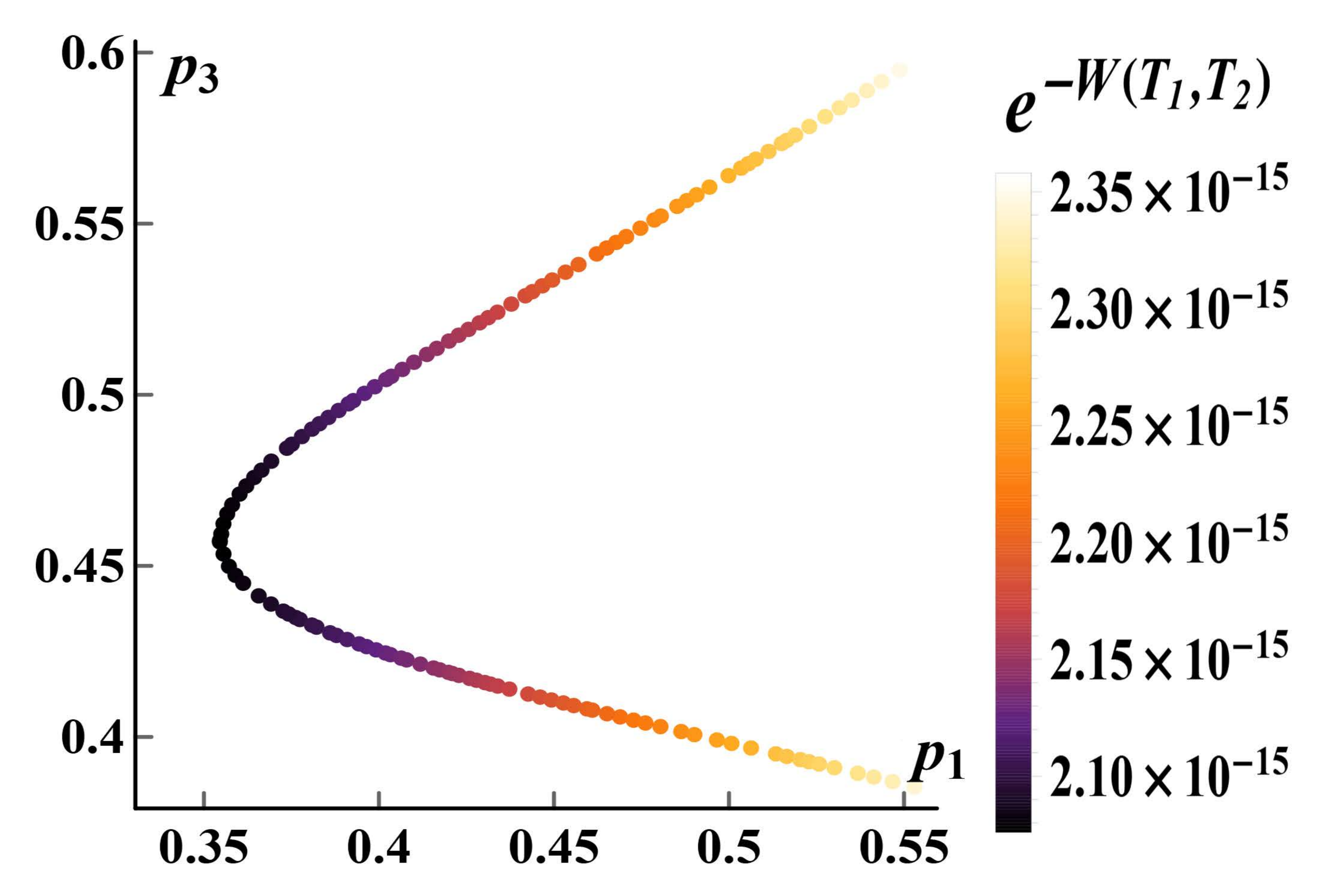}
\includegraphics[width=5.9 cm,height=3.1cm]{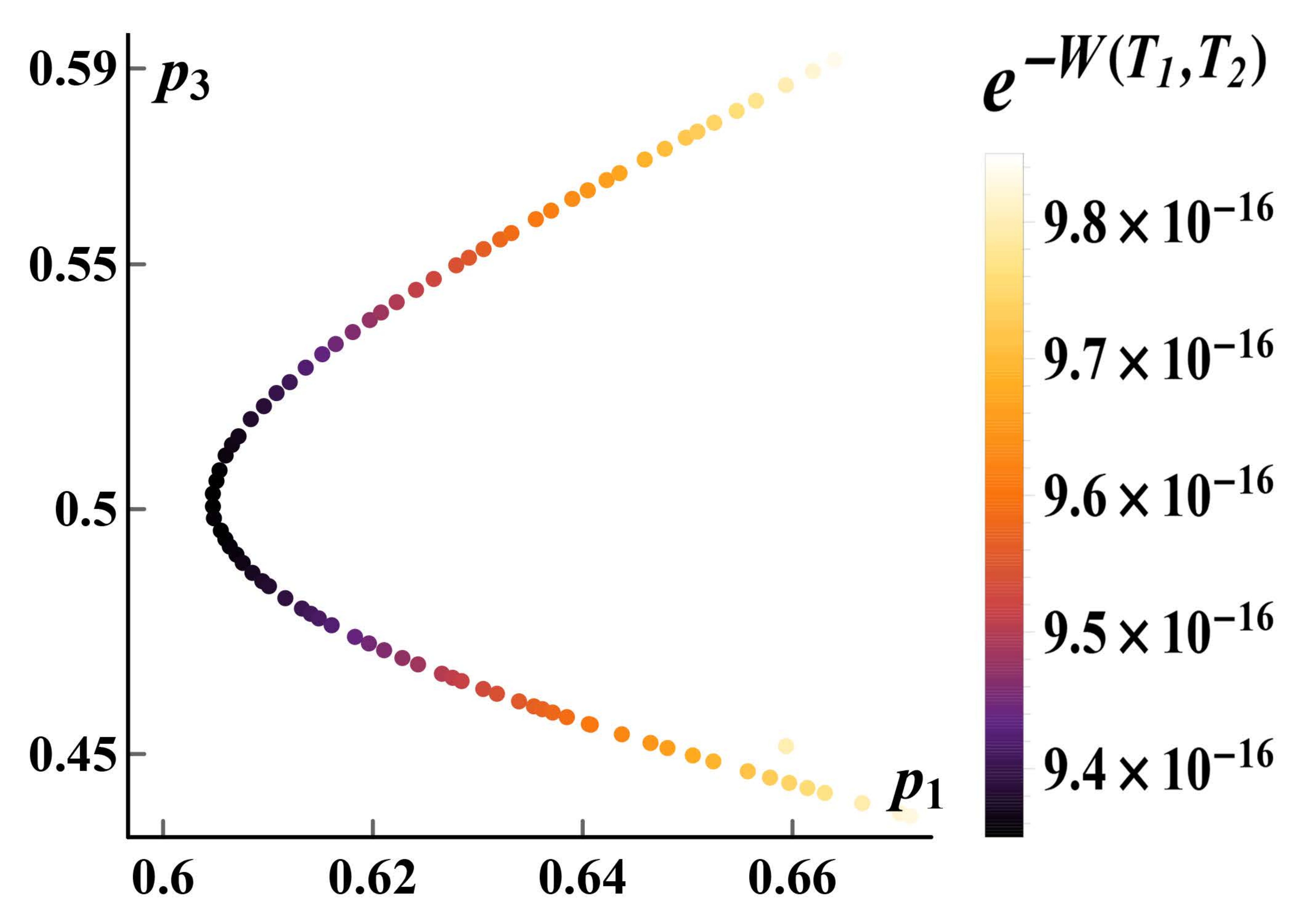}
\includegraphics[width=5.9 cm,height=3.1cm]{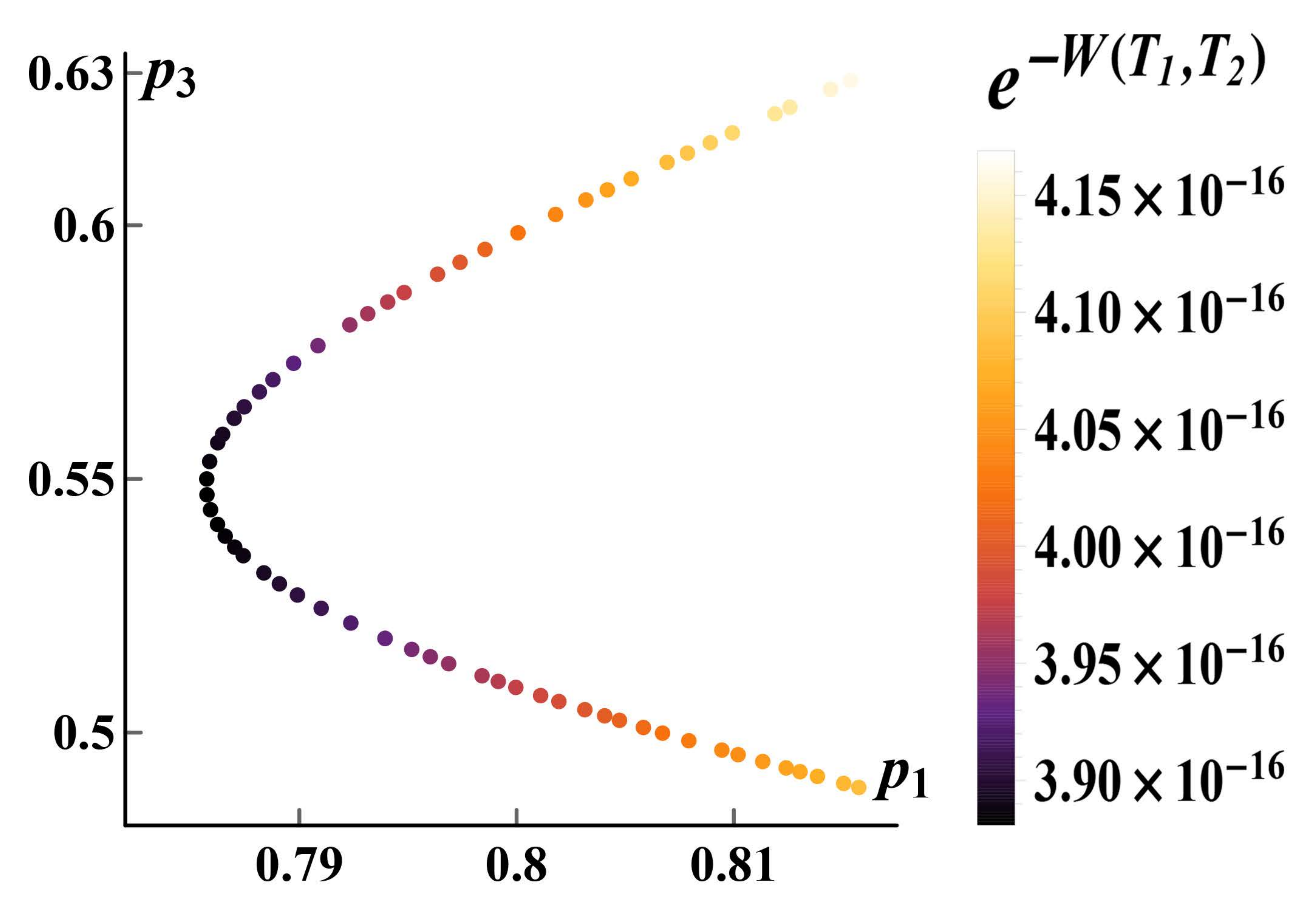}
\caption{Tunneling rates for each family of closed trajectories}
\label{f4}
\end{figure*}

Returning back to evaluation of   $W[T_1, \,\tilde {T}]$, we see that for a single closure the spinor trace yields $\cos{\pi}$, which can be absorbed into the prefactor.  This further simplifies imaginary part of the effective action to:
\begin{eqnarray}
\text{Im}\, \Gamma[A_{\mu}]\approx\sum_{\substack{\text{$p$}}}\, e^{-W[T_1, \,\tilde {T}]}, \quad W[T_1, \,\tilde {T}] > 0
\label{seffact2}
\end{eqnarray}

Having obtained the closed orbits,  decay rate can now be determined upon evaluation of the action. For convenience, we use the configuration space where $W[T_1, \,\tilde {T}]$ reads
\begin{eqnarray}
W[T_1, \,\tilde {T}]=- T_1/\tilde{T} \int_0^{\tilde{T}} (\dot{x}^2_0 - \dot{x}^2_3)\, du.
\label{char}
\end{eqnarray}
Figure \ref{f4} shows vacuum decay rate, i.e tunneling probability for each family of closed trajectories. Decay rate is consistently higher for the family of orbits with relatively lower momenta, but  it surprisingly increases along all the curves, until the invariant tori break. This behavior suggests that decay rate is maximized on the boundary of tori and it would be interesting to see whether this remains true for other cases. Comparing the results with  purely time-dependent counterpart of (\ref{field}), we see that pair production rate for the overlapping range of canonical momentum drops down by an order of magnitude \cite{gies1}. The fall of vacuum decay rate in the presence of a perpendicular magnetic field can in fact  be seen directly from the pair production formula for the uniform field. By using the peak values of  electric and  magnetic field in the invariant field strength, it is easy to see that uniform field approximation  also yields around an order of magnitude drop for the decay rate, with respect to the purely electric field case. This result can intuitively be understood by using the qualitative description of the virtual particles: the external electric field along $x_1$  separates the virtual electron-positron pairs, whereas the magnetic field applied in the perpendicular direction tries to bring these virtual pairs together.  

To summarize, worldline method relies on the existence of bounded motion. In this respect, integrability of the Hamiltonian plays an important role in the pair production process. Given that  $\mathcal{\hat{H}}$ is integrable, the use of multiperiodic trajectories in quantum tunneling naturally leads to the quantization of the fermionic modes by $\mathcal{\hat{H}}$ and $\mathcal{\hat{C}}$. For the external field given in (\ref{field}), multiperiodic trajectories  form a one parameter family  due to the existence of an additional constant of motion, $p_1$. But it is useful to keep in mind that tunneling may also occur via isolated trajectories.  To make a final remark on the use of transformation in (\ref{eff2}), we would like to point out that averaging methods similar in spirit have been used in atomic molecular physics to obtain the spectrum of multidimensional bound systems. For instance,  the evaluation of action by using the caustics  or the Poincar\'e sections of quasiperiodic trajectories has successfully generated the energy eigenvalues of multidimensional systems, such as 2D coupled harmonic oscillators \cite{noid1}. Here in our approach averaging arises naturally, with only requiring the periods of motion. 

The worldline picture provides a valuable semiclassical tool to calculate the nonperturbative decay amplitudes yet there are several noteworthy aspects of the problem that remain unaddressed. The first one is the prefactor contribution to the decay amplitude. A numerical method to obtain the prefactor and Maslov index  in one dimensional inhomogeneities was developed in \cite{dunne2}. Extension of such method to spatiotemporal backgrounds  is necessary for the completeness of the worldline approach and will be the subject of a future work.  Another issue is related to the constant of motion, $\mathcal{C}$. The determination of  $\mathcal{C}$ from the given orbit data would  be a desirable asset for the worldline method because the values that $\mathcal{C}$ takes, in conjunction with the topological index, can reveal a good deal of information about the spectrum of the created particles. Final and perhaps more interesting aspect is the quantum interference effects seen in the particle spectrum. Basic mechanism behind the interference phenomenon, in time-dependent fields for instance, can be understood in terms of  worldlines that are glued together at critical (conjugate) points on the complex time plane. Multiperiodic counterparts of such composite orbits are expected to come with a rich topological structure, but whether this basic procedure can be extended to the multidimensional cases remains to be seen.
\bigskip

I thank Gerald Dunne for discussions and acknowledge the support from T\"{U}B\.{I}TAK through the Grant 112C008. Part of the numerical computations reported in this work was performed at T\"{U}B\.{I}TAK ULAKB\.{I}M, High Performance and Grid Computing Center.


\begin{thebibliography}{999}
\bibitem{note}
The leading order approximation for the  current,  $\partial_{\nu} F_{1\nu} (\nu=0,3)$
can be given  by using the classical trajectories. For the closed tunneling orbits, where $\sigma \gg m^{-1}$,
we have $|\partial_{\nu} F_{1\nu}| \ll |F_{1\nu} \dot{x}_{\nu}|$. But if the external field gets localized down to the scale of vacuum fluctuations, magnitude of the current becomes at the order of field strength, and significantly contributes to the commutator, $[d\,\hat{x}_{1}/du,\, \not{\hspace*{-.1cm}\hat{D}^2}]$.
\end{thebibliography}
\end{document}